\documentclass[epjCONF,columns]{svjour} %for 2 columns format
\usepackage{graphics}
\usepackage[varg]{txfonts} % Times fonts
\usepackage[latin1]{inputenc}
\usepackage{xspace}
\usepackage[]{units}
\session-title{Hadron Collider Physics Symposium, 2011}

\newcommand{\ipb}{~pb\ensuremath{^{-1}}\xspace}
\newcommand{\ifb}{~fb\ensuremath{^{-1}}\xspace}
\newcommand{\met}{\mbox{\ensuremath{\slash\mkern-12mu{E}_{\mathrm{T}}}}\xspace}
\newcommand{\vmet}{\mbox{\ensuremath{\slash\mkern-12mu{\vec{E}}_{\mathrm{T}}}}\xspace}
\newcommand{\mht}{\mbox{\ensuremath{\slash\mkern-12mu{H}_{\mathrm{T}}}}\xspace}
\newcommand{\HT}{\ensuremath{H_{\mathrm{T}}}}
\newcommand{\vpt}{\ensuremath{\vec{p}_{\mathrm{T}}}}
\newcommand{\pt}{\ensuremath{p_{\mathrm{T}}}}
\newcommand{\tauHad}{\ensuremath{\tau_{\mathrm{h}}}}
\newcommand{\ttbar}{\ensuremath{t\bar{t}}}

\begin{document}
\title{Supersymmetry Searches at the Compact Muon Solenoid (CMS) Experiment, 2011}
\author{S. A. Koay\inst{1}\fnmsep\thanks{\email{sakoay@cern.ch}}, on behalf of the CMS Collaboration}
\institute{University of California, Santa Barbara}
\abstract{
The discovery/exclusion of Supersymmetric models for fundamental interactions of particles is one of the milestones targeted by the Large Hadron Collider (LHC), and in particular comprises of a large part of the physics program of the CMS experiment. Since the initial measurements using the 36\ipb of integrated luminosity delivered by the LHC in 2010, presently available results utilize about one fifth of the data delivered in 2011, i.e. in the ballpark of 1\ifb, significantly extending the world limits placed on gluino and squark production signals. An overview of these analyses is presented, highlighting four that had been newly made public as of the date of this conference. The evidence for Supersymmetry (SUSY) is still elusive, and a discussion follows as to where current searches have not yet probed, also pointing out where they might have difficulty ever probing without dedicated arrangements.} %end of abstract
\maketitle
\section{Introduction}
\label{intro}
The description and motivation of Supersymmetric extensions to the Standard Model (SM) of particle physics are summarized elsewhere in these Proceedings. The main experimental signature of such models is an apparent non-conservation of [transverse] momentum arising from the lightest SUSY particle (LSP) passing invisibly through the detector, as it is prohibited to decay if the new ``R-parity'' symmetry (required for proton stability) is conserved. As such, most SUSY searches feature use of the variable $\met$, defined as the magnitude of missing transverse momentum in the event, to discriminate between SUSY signal and SM backgrounds. Some analyses instead focus on variants of $\met$ used in conjunction with other kinematic information about the event, designed to provide superior background rejection, simplicity of background estimation techniques, and/or more direct correlation with signal mass scales. One of the major strengths of the CMS SUSY search program is in fact its diverse coverage of final states and features potentially sensitive to the yet-unknown nature of Beyond Standard Model (BSM) physics; a classification of analyses and their evolution from the 2010 to 2011 dataset is given in Section~\ref{sec:summary}. Sections~\ref{sec:jzb} to~\ref{sec:razor} go into a little more detail about four of the 2011 analyses, distinguished by their main search variables. SUSY exclusions are often hard to conclusively state due to the large number of theory unknowns and loopholes that allow otherwise feasible signatures to fall below experimental thresholds. With this caveat in mind, results are discussed in Section~\ref{sec:results} in terms of limits in a 2-parameter benchmark plane of the Constrained Minimal SUSY Model (CMSSM), as well as Simplified Model Spectra (SMS) for a more distilled, per-signal-process understanding. Section~\ref{sec:conclusion} concludes.

\section{SUSY searches from 36\ipb to 1\ifb}
\label{sec:summary}
\begin{figure}\begin{center}
\resizebox{1\columnwidth}{!}{\includegraphics{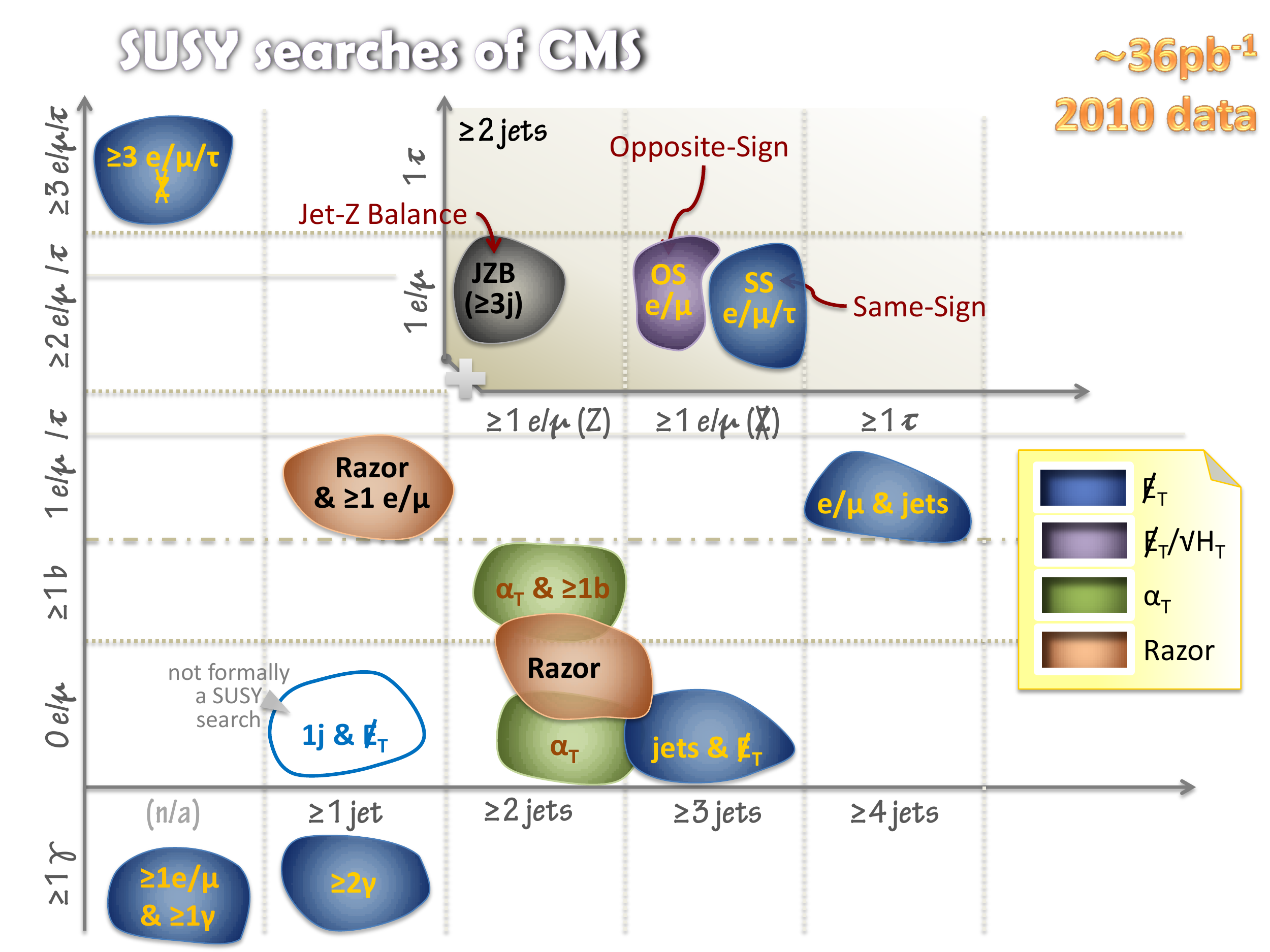} }
\caption{SUSY searches in 2010, by object multiplicity signatures.}
\label{fig:susy-2010}
\end{center}\end{figure}

\begin{figure}\begin{center}
\resizebox{1\columnwidth}{!}{\includegraphics{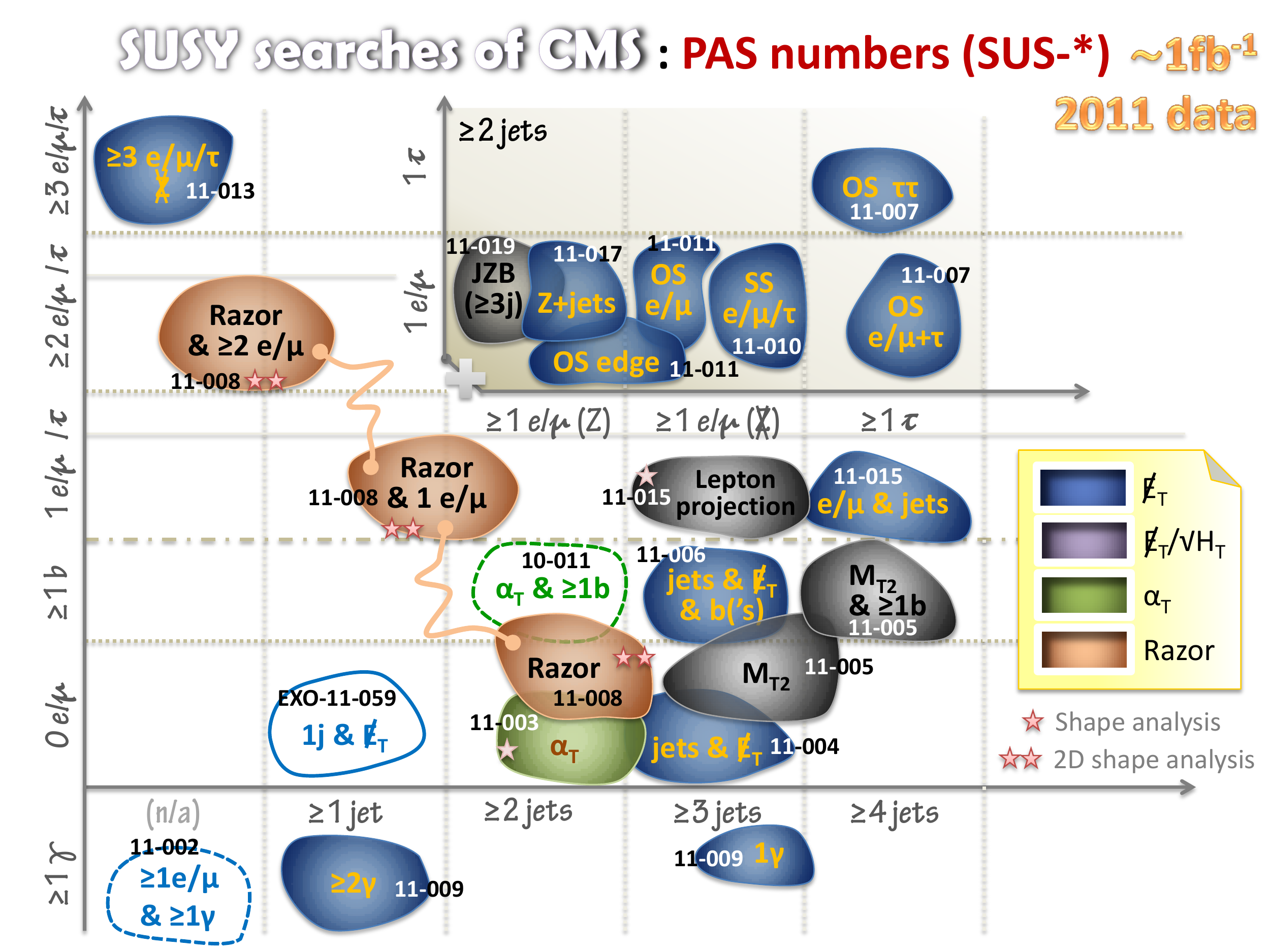} }
\caption{SUSY searches in 2011, by object multiplicity signatures.}
\label{fig:susy-2011}
\end{center}\end{figure}

One model-independent way of charting BSM territory is to list signatures with all possible multiplicities of final-state objects in the event. As of the time of this writing, it suffices to consider the following object types actively used in CMS SUSY searches: jets (inclusive), jets tagged as originating from b-quark decays, photons, and leptons. For the latter, a practical distinction is made between electrons and muons, versus tau leptons which are experimentally more difficult to identify. Figure~\ref{fig:susy-2010} shows the 2010 analyses in this view, with each analysis placed in the box corresponding to its loosest search region. The inset axes provide a ``zoom in'' of di-lepton signatures, which has a rich array of features that can be exploited based on correlations between the two leptons. These early searches are counting experiments focusing on basic signatures, mostly with jet multiplicity requirements as only sensitivity to the high-cross-section gluino and squark production modes is expected with 36\ipb of integrated luminosity. As seen in Figure~\ref{fig:susy-2011}, the 2011 searches expand significantly on this program in terms of number of examined final states, and demonstrates the CMS emphasis on cross-checks by looking at the same channels through different features. Moreover, techniques are evolving from cut-and-count to shape analyses of the most discriminating distributions, and correlating information from multiple channels (Section~\ref{sec:razor}).

\subsection{Jet-Z Balance (JZB)~\cite{jzb}}
\label{sec:jzb}
The main idea behind the JZB method is that in BSM events where a Z boson is produced in the decay chain, the angle between the boson and the jets in the event are decorrelated due to the non-visible presence of LSP's that make up the rest of the transverse momentum balance. This is in contrast to SM Z production where the boson recoils back-to-back with the jet system, up to jet resolution effects. The variable examined in this search is JZB$~\equiv |\sum_{\mathrm{recoil}} \vpt| - |\vpt(\ell^{+}\ell^{-})|$, where the first sum is over jets and the second the boson $\pt$ as reconstructed from a di-lepton (e, $\mu$) selection. The distribution of this variable is more or less symmetric about 0 and steeply falling for SM Z boson production events, which allows this background to be simply estimated by assuming that the number of events in the positive tail is equal to that in the negative tail. The other significant background in this search is SM top pair-production ($\ttbar$) events where both top quarks decay leptonically. This and other flavor-symmetric backgrounds are predicted as the average yield in 3 control regions: one requiring opposite-sign (OS) lepton pairs with invariant mass within a Z window, and the other two requiring OS or same-sign (SS) lepton pairs outside of the Z mass window. No significant excess is observed in data when accounting for the uncertainties of the prediction, although a slight tension is seen in the most extreme search region (JZB$ > $150~GeV).

\subsection{Opposite-sign $\tau$ pair~\cite{ditau}}
\label{sec:ostau}
Tau leptons are a challenging reconstruction task despite simple and well-known decay modes, due to a huge background from jets. About one third of taus decay leptonically into (soft) electrons and muons, and can be tagged with relative ease. The rest ($\tauHad$) decay into hadrons and neutrinos, which require a dedicated reconstruction algorithm including isolation cuts, and yields a selection efficiency of about 30-40\% here. Two different strategies are pursued in this search, one for the lower branching ratio but higher purity e/$\mu + \tauHad$ channel, and another for the higher branching ratio but more contaminated $\tauHad + \tauHad$ channel. In both cases an excess is looked for in region(s) of high $\HT$ (scalar sum momenta of jets) and $\met$ (or equivalently $\mht$, the vectorial sum $\pt$ of jets). The dominant background for the e/$\mu + \tauHad$ channel is $\ttbar$ production where both tops decay into tau leptons. The distribution of $\met$ in these SM events is highly similar to that where both tops decay into electrons or muons instead, modulo some correction factors and accounting for the different efficiencies of tau vs. electron/muon selection. The other substantial background are events with prompt sources of electrons/muons (e.g. from W boson production and decay), but a jet misidentified as a $\tauHad$. A sideband extrapolation method is used to estimate the tight-isolated population from a loose-isolated control sample. The $\tauHad + \tauHad$ channel on the other hand has several background sources, notably $\ttbar$, multijet, W and Z production, the latter in particular where the Z boson decays invisibly. The strategy here is to select using various cuts four control regions that are enriched in these SM processes, and translate to a prediction in the search region by accounting for their relative selection efficiencies. Overall, the $\tau$ pair search observes data in good agreement with prediction.

\subsection{Multi-leptons~\cite{multilepton}}
\label{sec:multilepton}
The multi-lepton search looks in regions with 3 or more leptons, where all three flavors of leptons (e, $\mu$, and $\tau$) are accepted, but at least one electron or muon is required to increase purity. The search regions are very clean, which allows the analysis to probe a wide range of kinematic regimes including regions without $\met$ requirements, as it is not required to regulate the SM background---this makes it also a golden channel for some signatures of R-parity-violating SUSY. Despite the apparent simplicity of this search however, careful work must be done in order to understand the backgrounds, as they originate from rare processes that may or may not be well-modeled by simulations. The non-prompt backgrounds are obtained from data-driven methods, for example the case of photon radiation off one of the leptons in Z boson production and decay, where the photon later converts (either internally or in the detector) and leads to a final state with an additional dilepton pair, therefore either promoting the event to a 3 or 4 lepton final state depending on how many of the conversion leptons are reconstructed. This background can be seen in the tri-lepton invariant mass distribution as a clear peak about the Z mass, which allows the photon conversion factor to be estimated in a low $\met$, low $\HT$ control region and later applied to predict the contamination rate in the search regions using $\ell^{+}\ell^{-}\gamma$ events. At the end, a detailed catalog of observations in the 3 and $\geq$ lepton channels have been published, covering possibly many BSM theory footprints, and all of which are in agreement with prediction.

\subsection{Razor~\cite{razor}}
\label{sec:razor}
The Razor analysis utilizes variables that are more directly related to the mass scale of potentially produced BSM particles, than the ``staple'' quantities $\met$ and $\HT$. This is achieved by considering the conceptually simple case where two BSM particles of mass $M_0$ are pair-produced at rest and each decay into a single jet and the LSP. In such a case the jet is known to have monochromatic energy $M_\Delta\equiv(M_0-M_\mathrm{LSP})^2/M_0^2$. The first step is thus to relate multi-object final states to this nominal system, by clustering the objects into two ``megajets'' via some hemisphering algorithm. The effect of the unknown $z$-boost of the system is then reduced by boosting it to a frame where the $z$-momenta of the two megajets are equal and opposite; the total energy in this frame is the first Razor variable, $M_R\equiv\sqrt{(E_1+E_2)^2 - (p_z^1+p_z^2)^2}$, and has a distribution that peaks about $M_\Delta$. Another sensitive variable is an average transverse mass computed by dividing $\met$ equally into two ``LSP momenta'', i.e. $M_T^R\equiv\sqrt{\nicefrac{1}{2}\left[\met (\pt^1 + \pt^2)-\vmet\cdot(\vpt^1+\vpt^2)\right]}$. The second Razor variable, $R \equiv M_T^R/M_R$, is strongly correlated with $M_R$ for SM background processes, for which the distribution peaks at small values in both $R$ and $M_R$, separating them from signal events for which the $R$ distribution is more spread out and $M_R$ is preferentially located around the (high) value of the produced BSM particle mass. Another large advantage of the Razor variables is that the SM distributions are seen to have exponential or sum-of-2-exponentials shapes in either one of the variables, with slopes that are controlled by cuts on the other variable. This allows for a very simple background estimation method, where each background is modeled by a 2D functional form in $R$ and $M_R$, and the initial parameters and constraints of which are extracted from control regions enriched in the respective background process. The analysis currently looks at data in six different channels: hadronic, single muon, single electron, and three di-lepton combinations (ee, $\mu$e, $\mu$$\mu$). A final combined fit for the final parameters and normalizations of these SM background shapes is performed in a low-$R$, low-$M_R$ sideband region of each of channel box, which extrapolates the yields into the search region according to the hypothesized functional form. The simplicity of this modeling permits a fairly sophisticated statistical analysis to be performed, correlating information from all six channels as well as performing a shape test of the 2D $(R,M_R)$ distribution in each box. Such a scheme has a clear advantage in being able to self-adapt to discovering signals in many varieties that reality may manifest. Sadly, good agreement between prediction and data is seen in all channels and regions in the search, with the lowest observation p-value being larger than 0.1.

\section{Results}
\label{sec:results}
\begin{figure}\begin{center}
\resizebox{1\columnwidth}{!}{\includegraphics{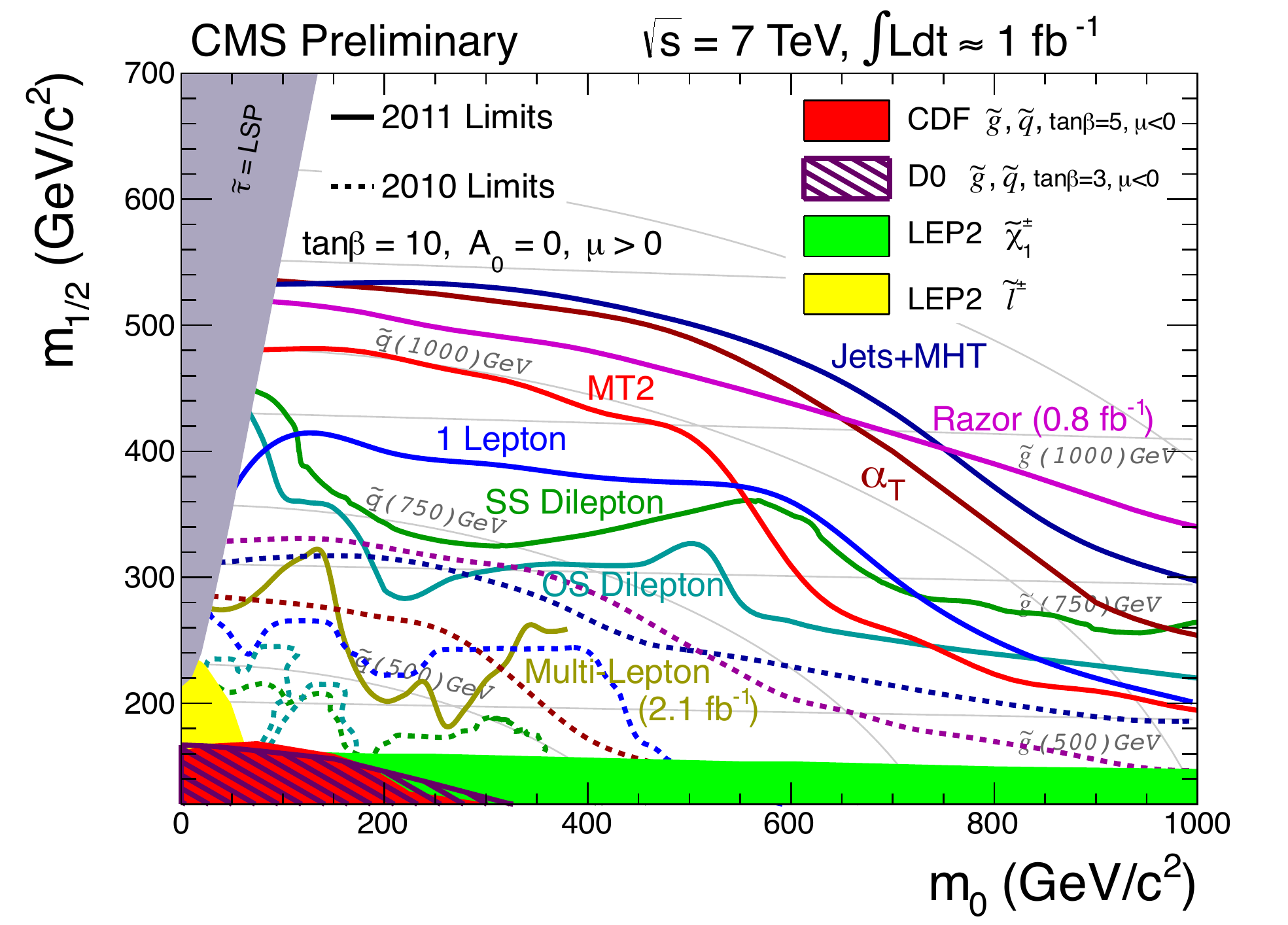} }
\caption{Observed limits from 2011 CMS SUSY searches in the CMSSM ($m_0$,$m_{\nicefrac{1}{2}}$) plane.}
\label{fig:cmssm-limits-2011}
\end{center}\end{figure}

With no excesses above expectation observed in data, limits have been set in several benchmark SUSY models, one of which is the CMSSM plane shown in Figure~\ref{fig:cmssm-limits-2011}. Contours of equal squark and gluino masses have been marked on this plane, from which it can be seen that when both species are active ($m_{\tilde{q}} \sim m_{\tilde{g}}$), a region up to about $m_{\tilde{q}} \lesssim 1100$GeV has been excluded by the most performant searches, corresponding to $m_{\tilde{g}} \lesssim 1200$GeV in the low $m_0$ region and $m_{\tilde{q}} \lesssim 900$GeV at higher $m_0$. When the squarks are decoupled however, i.e. the extreme $m_0$ region beyond the right edge of this plot, the exclusion reach peters out to $m_{\nicefrac{1}{2}} \lesssim 275$GeV ($m_{\tilde{g}} \lesssim 700$GeV). This large exclusion range may seem to threaten the feasibility of SUSY---or at least the CMSSM---as a description of nature, it should have mass scales not too far above the electroweak (EWK) scale of around 246GeV in order to achieve ``natural'' (non fine-tuned) EWK symmetry breaking. It has been noted~\cite{natural-susy-endures} however that most of the superpartners provide by-and-large irrelevant loop corrections to the Higgs mass, the only important ones being (a) two stops and one sbottom below 700GeV; (b) two higgsinos below 350GeV; and (c) a gluino below 1.5TeV. Due to the much smaller cross-sections, present searches have little reach for (a) and (b), and are still far from the 1.5TeV gluino mass range depending on kinematic details of the gluino decays.

\begin{figure}\begin{center}
\resizebox{1\columnwidth}{!}{\includegraphics{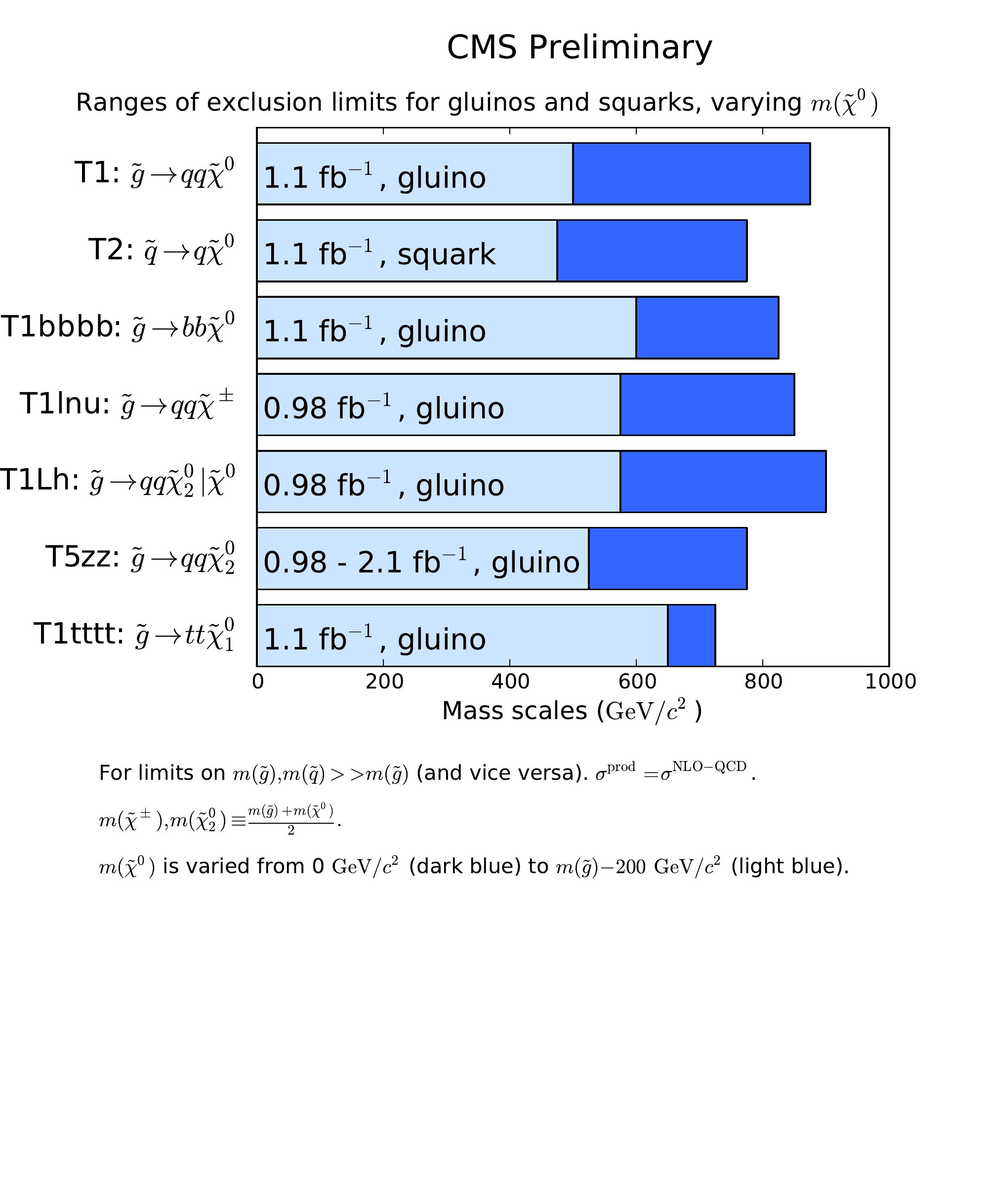} }
\caption{Observed limits from 2011 CMS SUSY searches in several SMS models, with the stated caveats.}
\label{fig:sms-limits-2011}
\end{center}\end{figure}

As it is often hard to extrapolate from limits set in particular models like the CMSSM to other models with different types and/or relative cross-sections of processes, both CMS and ATLAS collaborations complement such limits with those set in Simplified Model Spectra (SMS) benchmarks. Each SMS consists of a small list of new particles and their decays, most of the time implying one or  very few production-and-decay topologies. They can be thought of as building blocks of more complete models, i.e. effective theories, and limits are set assuming that these reactions occur in isolation (100\% branching ratio). One consequence of this is that signal contamination is accounted for as applicable for the search in question, but only that from the particular SMS under study. A summary of SMS model exclusions using the 2011 data is shown in Figure~\ref{fig:sms-limits-2011}, where the best limits have been taken over all participating searches. It is seen that gluino and squark masses of up to 450-90GeV have been ruled out, contingent on the decay chains and masses of the produced colored new particles, and assuming that the difference between the produced and LSP masses is no smaller than 200GeV.

\begin{figure}\begin{center}
\resizebox{1\columnwidth}{!}{\includegraphics{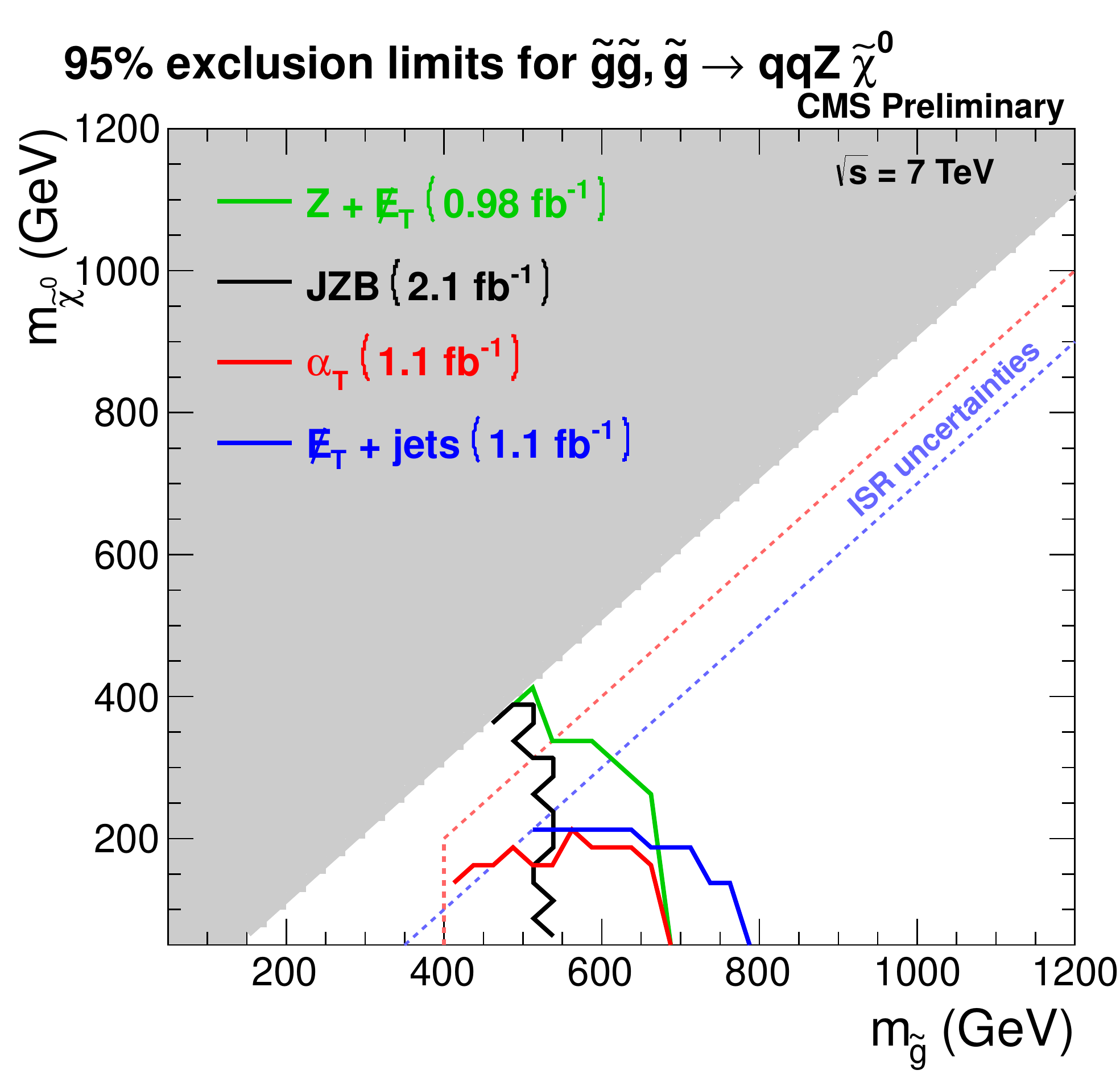} }
\caption{Observed limits from 2011 CMS SUSY searches in a $\tilde{g}\tilde{g}$ production SMS topology, where the gluino decays into two jets, a Z boson, and the LSP.}
\label{fig:T5zz-limits-2011}
\end{center}\end{figure}

Beyond being an alternate representation of results however, SMS models provide a physically intuitive understanding of how search performances depend on kinematic features and final state observables of various SUSY-like processes. Figure~\ref{fig:T5zz-limits-2011} is an example of exclusion limits set by four CMS SUSY searches in a $\tilde{g}\tilde{g}$ production SMS topology with four jets, two Z bosons, and two LSP's ($\tilde{\chi}^0$) in the final state. In the region where the mass spectrum is somewhat ``squeezed'', i.e. $m_{\tilde{g}} \sim m_{\tilde{\chi}^0}$, kinematic observables (including $\met$) become much softer, leading to a loss of exclusion from the otherwise most performant hadronic searches, $\alpha_{\mathrm{T}}$ and $\met+$jets, as they require hard cuts to improve purity. The importance of leptonic searches, here Z$+\met$ and JZB, are clearly seen to provide a complementary boost in sensitivity as their cleaner search regions allow relaxation of kinematic cuts. 

\section{Conclusion}
\label{sec:conclusion}

It has been the author's pleasure to present the progress of CMS SUSY searches in 2011 at this conference, which is a comprehensive program covering an increasingly large array of final states and features as increasing integrated luminosity and time allows. In particular, four of these searches---looking at the JZB, di-$\tau$, multi-lepton, and Razor variables---have just been approved for public presentation at this time. None of the searches so far have observed significant deviations from prediction in more or less 1\ifb of collision data, and have therefore set exclusion limits of up to more than 1TeV in squark mass and 700-900GeV in gluino mass in the CMSSM benchmark model, as well as limits in the range of 450-900GeV in squark and gluino masses for several single-topology simplified models.

An overall understanding of SUSY model configurations not yet excluded as of the time of this writing is as follows:
\begin{itemize}
\item A large portion of the CMSSM ($m_0$,$m_{\nicefrac{1}{2}}$) plane where $m_{\tilde{g}} \lesssim 1.5$TeV.
\item Models with squeezed mass spectra. Such signals require significant initial-state radiation to have appreciable $\met$ or $\HT$, which makes them very experimentally challenging to separate from background. This also leads to them suffering from large theory modeling uncertainties, which further degrades what may be said about their presence or lack thereof in data.
\item Models with lower than expected signal yields, which may for example occur if the production cross-section is lower than vanilla SUSY assumptions, or if there is significant branching ratio to other, less detectable final states.
\item Direct production of stops/sbottoms, due to their small cross-section. At low stop/sbottom masses where the cross-section is high, the similarity of signatures (in some decay channels) to SM $t\bar{t}$ production also makes it experimentally challenging to detect.
\end{itemize}

In addition, there are signatures that have not yet been explicitly looked at by CMS SUSY searches, including those with: (a) very high jet multiplicities; (b) very low jet multiplicities (e.g. if colored states are not accessible); (c) more complex object types like reconstructing top quarks in the final state; and (d) many more combinations of types and multiplicities of final state objects.

%However, stop/sbottom production via gluino decays have been ruled out to some degree, see Figure~\ref{fig:T2tt-limits-2011}.
%\begin{figure}\begin{center}
%\resizebox{1\columnwidth}{!}{\includegraphics{gluinoStopResult.pdf} }
%\caption{Observed limits from two 2011 CMS SUSY searches in a $\tilde{g}\tilde{g}$ production SMS topology, where the gluino decays into two top quarks and the LSP.}
%\label{fig:T2tt-limits-2011}
%\end{center}\end{figure}

\end{document}

% end of file template.tex